# Micro(point)contact spectroscopy of dilute magnetic (Kondo) alloys CuMn and CuFe

Yu. G. Naidyuk, O. I. Shklyarevskii, and I. K. Yanson

*Physicotechnical Institute of Low Temperatures, Academy of Sciences of the Ukrainian SSR, Khar'kov*

The method of microcontact spectroscopy is used to study alloys with magnetic impurities CuMn and CuFe in the range of concentrations 0.01-1 at.%. Minima or maxima (so-called zero-bias anomalies) were observed in the microcontact spectra of these alloys at voltages ~ 1 mV. The Lande *g*-factor for the Mn impurity in a Cu matrix was determined from the splitting of the minimum of the zero-bias anomaly in a magnetic field. The quantitative calculations carried out agree well with theory and permit determining both the important geometric contact parameter <K> i.e., the averaged geometric form factor, and the characteristics of the alloy itself, for example, $J/\varepsilon_F$, i.e., the ratio of the magnitude of the exchange interaction energy between an electron and the magnetic impurity to the Fermi energy, from the microcontact spectra.

The investigation of nonlinear effects in the electrical conductivity of point contacts is at the present time successfully used for studying the electron-phonon interaction in metals and some alloys.[1-4] The second derivative of the current-voltage characteristic $d^2I/dV^2(V)$ of such a microcontact or the microcontact spectrum at helium temperatures is proportional, in the range of energies eV from 0 to $k_B\Theta$ ($\Theta$ is the Debye temperature), to the electron-phonon interaction function.[2] However, the microcontact spectra exhibit features both at energies greater than $k_B\Theta$ and for eV « $k_B\Theta$ that are absent in the theory developed in Ref. 2. The investigation of such features is of interest in itself, since it greatly increases the range of applicability of microcontact spectroscopy for studying the interaction of conduction electrons not only with the phonon system of the metal, but also with other Bose excitations in conductors, as well as for studying scattering of electrons by impurities and defects.

In studying the electron-phonon interaction with the use of microcontact spectroscopy, the appearance of anomalies for eV « $k_B\Theta$ was usually associated with an imperfection of the region of contact, the presence within it of impurities or regions of strongly deformed metal. In Ref. 3, while studying the alloys CuNi and CuFe (~l at.% Fe), it was discovered that the presence of impurities in the contact region with localized magnetic moment (Fe and Cu) leads to the appearance of a deep minimum $d^2V/dI^2(V) < 0$ on the microcontact spectra at V ~ 1 mV, which corresponds to the maximum of dV/dI at V =0.
A theoretical analysis of the problem of current flow in microcontacts in the presence of magnetic impurities[5], showed that on $d^2I/dV^2$ (or $d^2V/dI^2$) there must be singularities at small energies and, in addition, the sign of the second derivative depends on the sign of the constant J, entering into the Hamiltonian of the exchange interaction of the electron with the magnetic moment of the impurity. In Refs. 6 and 7, using the alloys AuMn, CuFe, and CuMn (< 1 at.% Fe, Mn),it was shown that such an anomaly is indeed related to scattering of electrons by magnetic impurities in the contact region (Kondo effect).

In this paper, we carry out a detailed analysis of the alloys CuMn and CuFe in the range of concentrations 0.01-1 at.% Mn and Fe. Aside from a qualitative comparison of the experimental data with the results in Ref. 5, quantitative calculations were performed, which show good agreement between theory and experiment and permit determining a number of such important parameters as the average geometric form factors of the microcontact, as well as the ratio $J/\varepsilon_F$ ($\varepsilon_F$ is the Fermi energy).

EXPERIMENTAL PROCEDURE

We obtained the alloys studied by melting the starting 99.997-99.999% pure metals in a vacuum of ~ $10^{-6}$ torr at T - 1000°C for 5-8 h. The alloy CuFe was quenched from T ~ 900°C in liquid nitrogen, which decreased the probability of the formation of clusters of Fe atoms. We established the final concentration of Mn and Fe with the help of a spectral analysis. For all specimens, we measured the ratios $R_{293}/R_{4.3}$, which for the alloys CuMn agree with the results in Ref. **8** with a spread of not more than 10% for the corresponding concentrations. The characteristics of the alloys are presented in Table I. The preparation of specimens with the required geometry and the technique for measuring the second derivatives are described in details in Refs. 1 and 4. To this, it is only necessary to add that in order to remove the mechanical cold working and the deformed surface after mechanical working of the specimens, for CuMn (0.01 and 0.07% Mn), we used chemical etching in $HNO_3$, and for the remaining alloys we used electropolishing in $H_3PO_4$. For measurements in a magnetic field created by a superconducting solenoid, we used a microdisplacement system made of nonmagnetic materials. We note that the microcontact spectra of alloys with a Mn and Fe content < 0.1% can be affected by impurities accidentally entering into the contact region, so that for small concentrations about 100 spectra were obtained and were analyzed statistically. For large concentration, from 20 to 50 spectra were measured for each alloy. We performed the measurements at a temperature 1.7-4.2 °K in a magnetic field varying from 0 to 60 kOe. We used a computer to calculate the integral parameters of microcontact spectra.

TABLE I. Characteristics of Alloys Studied

| Alloy | c, at.% | $R_{293}/R_{4.2}$ | Impurities |
|---|---|---|---|
| CuMn | 0.6 | 1.53 ± 0.02 | Fe 0.003% |
| | 0.12 | 5.7 ± 0.1 | |
| | 0.07 | 9.5 ± 0.3 | |
| | 0.012 | 38 ± 0.2 | |
| CuFe | 0.25 | 1.41 ± 0.02 | Mn 0.004% |
| | 0.14 | 2.2 ± 0.1 | |
| | 0.01 | 11.2 ± 0.1 | |

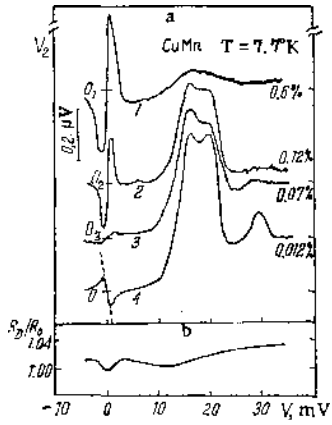

FIG. 1. Microcontact spectra of the alloy CuMn (a) and the behavior of the differential resistance for spectrum 1 (b). Parameters of the contacts: 1) $R = 2.4\ \Omega$., $V_{1,0} = 0.4$ mV; 2) $R = 2.2\ \Omega$., $V_{1,0} = 0.7$mV; 3) $R = 1.45\ \Omega$., $V_{1,0} = 0.65$ mV; 4) $R = 1.55\ \Omega$., $V_{1,0} = 0.35$ mV.

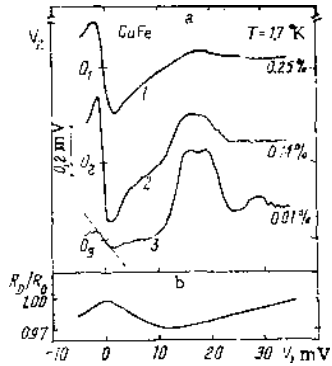

FIG. 2. Microcontact spectra of the alloy CuFe (a) and the behavior of the differential resistance for spectrum 1 (b). The parameters of the contacts are: 1) $R=5\ \Omega$., $V_{1,0} = 0.7$ mV; 2) $R = 1.8\ \Omega$., $V_{1,0} = 0.32$ mV; 3) $R = 1\ \Omega$., $V_{1,0} = 0.26$ mV.

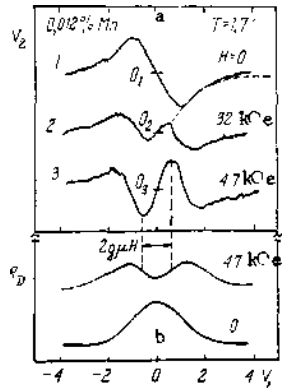

FIG. 3. The splitting of the negative anomaly in the alloy CuMn (0.012%) as a function of the magnetic field (a) and the behavior of the differential resistance for curves 1 and 3 (b). The parameters of the contact are: $R = 0.6\ \Omega$, $V_{1,0}=0.6$mV. The dashed line shows the $V^{-1}$ dependence.

EXPERIMENTAL RESULTS

Typical microcontact spectra for the alloys CuMn and CuFe are shown in Figs, la and 2a as functions of the concentration. It is evident that the presence of magnetic impurities leads to two very noticeable effects: appearance of "zero-bias anomalies" and suppression of electron- phonon interaction spectra in copper with increasing concentration. We note that the minimum on the microcontact spectrum (negative anomaly) corresponds to a maximum in the difference resistance $R_d = dV/dI$ at $V = 0$ (Fig. 2b), while the maximum (positive anomaly) corresponds to a minimum in $R_d$ (Fig. lb); in addition, for the alloy CuMn, a transition is observed from the negative anomaly to the positive anomaly with increasing Mn content. In Ref. 6, this change in microcontact spectra of AuMn was related to the effect of the internal magnetic field, increasing with concentration of the magnetic impurity. Indeed, in the alloy CuMn (0.012% Mn), for which the anomaly has a negative sign in the absence of a field, a transition is observed into the positive anomaly at $V_2(V) \sim d^2V/dI^2$, as a result of the splitting of the initial singularity in the external field (Fig. 3a). For the chara-cteristic $R_d(V)$, the effect of the field leads to the appearance of a trough at $V = 0$ (Fig. 3b). An estimate of the internal field of the alloy CuMn is made in Ref. 8, where $\langle H^2 \rangle^{1/2} \approx 150$ kOe c(%), i.e., for $c \sim 0.1\%$, H corresponds in order of magnitude to the external field, in which the splitting is observed. From here, it is understandable why the external field has no effect on the anomaly for c > 0.2-0.5% in CuMn and CuFe, since at such concentrations, $H_{in} \gg H_{out}$. Estimates of the internal field in CuFe are presented in Ref. 9:

$$H_{in}^{CuFe} \approx \begin{array}{l} 160\ \text{kOe}\ c\quad \text{for}\ c < 0.35\%; \\ 100\ \text{kOe}\ c^{1/2}\ \text{for}\ c > 0.35\%. \end{array}$$

Let us examine qualitatively the processes that lead to the appearance of such anomalies in microcontact spectra. In measuring the temperature dependence of the electrical resistance, the energy of electrons interacting with the impurity can vary by an amount $\sim kT$. On the other hand, the energy of electrons scattered in the microcontact changes by $eV$, i.e., the energy scale of the microcontact spectra can be compared to some effective temperature scale proportional to it (for example, for the thermal regime, $eV = 3.63\ kT$ at the center of the contact[10]). For this reason, an increase in the differential resistance for $V \rightarrow 0$ (negative anomaly $V_2 < 0$) corresponds to the usual Kondo increase in resistance with $T \rightarrow 0$ in the case $J < 0$.[11] With an increase in the concentration of the magnetic impurity, the internal field that arises splits the Zeeman sublevels of the magnetic ion and for a certain value of V the increase in resistance stops, since the energy of the electron is insufficient for scattering by the impurity accompanied by spin flipping due to the existence of an energy gap $g\mu H$.

We shall consider in greater detail the splitting of the negative anomaly in the alloy CuMn (~ 0.012% Mn). A similar phenomenon was also observed in the tunneling effect, where in order to explain the maxima in the conductivity at V =0, Applebaum[12] also invoked the Kondo effect. He predicted the appearance of a trough in the conductivity in the magnetic field with a width $2g\mu H$. According to Ref. 12, we can measure the quantity $2g\mu H$ for splitting of a negative anomaly in the microcontact spectrum (see Fig. 3) and determine the Lande g factor. A simple physical meaning can be ascribed to the measurement of $2g\mu H$ between the peaks of the second derivative for microcontact spectra. The Zeeman splitting arising in the magnetic field can be viewed as the appearance of a new branch in the spectrum of the interaction of conduction

TABLE II. Parameters of Microcontact Spectra, according to which $\langle K \rangle$ was calculated for CuMn and $J/\varepsilon_F$ for CuFe.

| Specimen No. | Alloy | c.at. % | R, Ω | $V_{1,0}$, mV | $V_2$ ($V = 2$ mV) μV | $\langle K \rangle \cdot 10^3$ | $\tilde{n}/n$ | λ | $J/\varepsilon_F$ |
|---|---|---|---|---|---|---|---|---|---|
| 1 | CuMn | 0.012 | 0.8 | 0.5 | 0.040 | 82 | 18 | 0.22 ± 0.04 | 0.0815 [7] |
| 2 |  | 1.55 | 0.35 | 0.022 | 210 | 43 | 0.20 ± 0.04 |  |  |
| 3 |  | 0.01 | 1 | 0.26 | 0.028 | 160 | 31 |  | 0.13 |
| 4 | CuFe | 0.01 | 0.5 | 0.29 | 0.049 | 110 | 20 | 0.19 ± 0.04 [14] | 0.15 |
| 5 |  | 0.14 | 1.8 | 0.32 | 0.262 | 120 | 22 |  | 0.12 |
| 6 |  | 0.14 | 4.2 | 0.49 | 0.315 | 130 | 25 |  | 0.11 |

Note. Specimens 3 and 5 correspond to spectra 3 and 2 in Fig. 2a; specimen 2 corresponds to spectrum 4 in Fig. la.

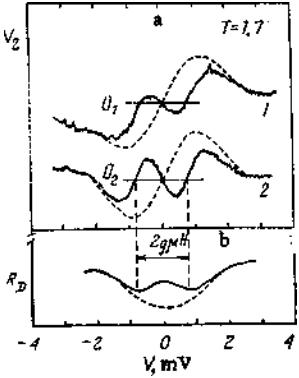

FIG. 4. The splitting of the positive anomalies in a magnetic field (a) and behavior of the differential resistance for curve 2 (b): 1) the alloy CuMn (0.07%), R = 3.7 Ω, $V_{1,0}$ = 0.5 mV; 2) the alloy CuFe (0.01%), R = 1 Ω, $V_{1,0}$ = 0.43 mV; --------) H=0; —) H = 60 kOe.

electrons with the magnetic moment of the impurity, which is what is reflected in the microcontact spectrum as a peak, similar to the phonon branches in the case of the Einstein spectrum. The measurements of 2gμH performed in this manner for 12 spectra gave the value g = 1.93 ± 0.07, which agrees with the data in Ref. 13, where g = 2 for Mn in a copper matrix. This is another convincing proof for the magnetic nature of such anomalies.

We observed such splitting of several positive anomalies (Fig. 4), arising in separate microcontact spectra of CuMn with a low Mn (< 0.12%) concentration and in the alloy CuFe (0.01%), which, apparently, are related to the increase in the content of the magnetic impurity (probably, Mn) near the contact and with the presence of an interaction between them. For such anomalies, the concentration of the magnetic impurity near the contact remains unknown and, therefore, the magnitude of the internal magnetic field also remains unknown. In this case, the value of 2gμH assuming that g = 2 corresponds to the splitting between peaks in $R_d$ or zeros of the second derivative (Fig. 4).

There is another series of experimental facts that does not as yet have a convincing explanation. These include the constancy of the sign of the anomaly $V_2(V) < 0$ in the alloy CuFe with increasing Fe content and the absence of splitting of this negative anomaly in a magnetic field even for small Fe concentrations. We note that the alloy CuFe differs considerably from the alloy CuMn by the fact that, first of all, iron does not form a solid solution in a Cu matrix and the Fe atoms form into clusters and, second, the temperature dependence of the electrical conductivity of these alloys at T ~ 1°K, due to the magnetic impurities, differs considerably.[8,9]

## COMPARISON OF THE OME L'YANCHUK-TULUZOV THEORY WITH EXPERIMENT

According to Ref. 5, the second derivative of the current-voltage characteristic of a microcontact, when scattering by impurities with a localized magnetic moment is taken into account, equals

$$\frac{d^2I}{dV^2} = -\frac{1}{R_1} J \frac{N(0)}{n} \begin{cases} 2/V, & eV \gg kT, \\ e^2V/(kT)^2, & eV \ll kT, \end{cases} \quad (1)$$

where

$$\frac{1}{R_1} = \frac{1}{R_0} \frac{J^2 S(S+1)}{2\pi\hbar^4} \frac{c}{n} \frac{\oint \frac{dS_p}{v_p} v_z \oint \frac{dS_{p'}}{v_{p'}} \tau_0(p,p')}{\oint (dS_p/v_p) \oint (dS_{p'}/v_{p'})}; \quad (2)$$

$R_0$ is the resistance of a clean contact; c is the impurity concentration; n is the electron density; S is the spin of the impurity; N(0) is the electron density of states per spin; e is the electron charge; $\tau_0$(p, p') is the characteristic relaxation time, determined by the shape and dimensions of the contact. The integration is carried out over the Fermi surface. According to Ref. 2, for a circular opening with diameter d, we have $\tau_0(p,p') = 4dK(v, v')/3\pi v_z$, where K(v, v') is the geometrical form factor entering into the definition of the microcontact electron-phonon interaction functions G(V) ~ $d^2I/dV^2$.[1,2] Using the free-electron approximation, $R_1$ can be expressed in terms of the average, over the Fermi surface, value of K(v, v) (see Eq. (19) in Ref. 1), while the magnitude of the second derivative $d^2I/dV^2$ can be expressed in terms of experimentally measured quantities. In this case, Eq. (1) reduces to the form

$$v_2 = 3\sqrt{2} V_{1,0}^2 k_F dc S(S+1) \langle K \rangle \left(\frac{J}{\varepsilon_F}\right)^3 \begin{cases} 2/V, & eV \gg kT; \\ (e/kT)^2 V, & eV \ll kT. \end{cases}$$

Here $k_F$ - and $\varepsilon_F$ - are the Fermi momentum and energy; $V_{1,0}$ and $V_2$ are the effective values of the first and second harmonics of the modulating signal.

We note first the qualitative agreement between theory and experiment. Indeed, for V→0 (eV ≪ kT), the quantity $V_2$ varies linearly with V (Figs. 1 and 2, curves 4 and 3). As the temperature decreased, an increase was also observed in the slope of the microcontact spectra at V = 0. For energies eV > kT, the variation in $V_2$ follows the dependence $V^{-1}$; which is also observed on the experimental curves for CuFe (Fig. 2a) and CuMn (Fig. 3a, curve 1) in the region 1-3 mV (this dependence is discussed in greater detail below). From Eq. (3), we can obtain an expression for <K>, for example, in the case eV » kT, for the alloy CuMn:

$$<K> = -18.3 V_2 V / V_{1,0}^2 \, d \, c. \qquad (4)$$

The following values were used for the constants in Eq. (4): $S = 5/2$. $J/\varepsilon_F = 0.0815$ (Ref. 8), $k_F = 1.36 \cdot 10^8$ cm$^{-1}$. In addition, $V_2$ was measured in microvolts, $V_{1,0}$ in millivolts, and d in angstroms (according to Ref. 1, $d = 300/R^{1/2}$ Å for a clean opening, if R is measured in ohms). Since we used the asymptotic equations (3) for eV » kT, the value of $V_2(V)$ from the microcontact spectrum must be taken after the minimum on the curve. However, for V > 3 mV, the effect of the electron-phonon interaction spectrum of copper becomes noticeable, so that we choose the value $V_2$ for 4-5 points in the region 1.2-2.5 mV. In this case, the spread in the values of <K>, calculated according to Eq. (4), did not exceed 20%, which indicates that the curve $V_2(V)$ corresponds well to the law $V^{-1}$ in this interval. The values of <K> calculated in this manner were averaged; their values for two spectra are presented in Table II. For the same specimens, the parameters $\lambda` = 2 \int G(V) V^{-1} dV$ can be calculated from the microcontact spectra of the electron-phonon interaction in copper as proposed in Ref. 1, as well as the fundamental electron-phonon interaction constant $\lambda = <K> \lambda`$ These values are also presented in Table II. It is evident that the quantities $\lambda$, calculated in this manner with the use of <K>, agree well with one another and with the data in Ref. 14, where $\lambda = 0.19 \pm 0.04$.

We determined the ratio $J/\varepsilon_F$ for the alloy CuFe using Eq. (3) and assuming that $<K> = \lambda`/\lambda$ is known. The values of $\lambda`$ were determined from the electron-phonon interaction spectra for specimens with known Fe concentration; the quantity $\lambda = 0.19 \pm 0.04$ was taken from Ref. 14; and for iron it was assumed that $S = 2$. The calculations carried out for several CuFe alloys with concentrations 0.14 and 0.01% gave the ratio $J/\varepsilon_F = 0.13 \pm 0.02$ (see Table II). This agrees well with the data in Ref. 15, where the quantity J for CuFe equals $0.91 \pm 0.2$ eV, which leads to the value $J/\varepsilon_F = 0.13 \pm 0.03$, if we take $\varepsilon_F = 7$ eV for Cu from the free-electron approximation. This value for $J/\varepsilon_F$ agrees with the fact that the introduction of a Fe impurity increases the specific resistance of the alloy CuFe, related to scattering by magnetic impurities, several times greater than in the alloy CuMn with the same Mn content (see Refs. 8 and 9). We note that for low impurity concentrations, when the elastic mean free path of the electron $l_i >> d$, as shown in Ref. 1, $<K> = 1/4$ and the ratio $J/\varepsilon_F$ can be determined without calculating $\lambda`$.

Thus, the basic results of this work reduce to the following. The microcontact spectra of CuMn and CuFe were measured in the range of concentrations 0.01-1 at.%, in which zero anomalies were observed (minima or maxima) at voltages ~ 1 mV. The measurements of the splitting of the negative anomalies in a magnetic field for low concentrations 0.01% Mn permitted determining the Lande g factor. The splitting of separate positive anomalies was observed.

The quantitative calculations performed, according to the theory in Ref. 5, permit obtaining <K>, i.e., finding experimentally the geometric characteristics of the contact entering into the expression for $R_1$. This is especially important from the point of view of microcontact spectroscopy, since <K> is used to calculate the integral parameters of the electron-phonon interaction from microcontact spectra and the given method makes it possible to determine it independently. On the other hand, as shown for the example of CuFe, it is possible to determine from the microcontact spectra the quantity $J/\varepsilon_F$, which is important in studying kinetic phenomena in magnetic alloys. Probably, this method is simpler than precision measurements of the temperature dependence of the specific resistance or, at least, it represents a new independent method for determining $J/\varepsilon_F$.

In conclusion, we thank I. G. Tuluzov for consultation concerning Ref. 5, L. E. Usenko, who carried out the spectral analysis of the alloys, A. A. Lyshkh and N. N. Gribov for help in carrying out the experiments and preparing the alloys, and V. Demirskii, who provided the CuMn (0.6 and 1 at.%) alloys.